\documentclass[runningheads]{llncs}
\usepackage{graphicx}
\usepackage{booktabs}
\usepackage[para,online,flushleft]{threeparttable}
\usepackage{makecell}
\usepackage{amsmath}
\usepackage{amsfonts} 

\RequirePackage{algorithm} % for algorithms
\RequirePackage{algorithmic}
\DeclareMathOperator*{\argmin}{argmin}

\begin{document}
\title{Autofocusing+: Noise-Resilient Motion Correction in Magnetic Resonance Imaging}
% ONE OF THE MAIN QUESTION: artifact or artifact?

\titlerunning{Autofocusing+: Noise-Resilient Motion Correction in MRI}
% \author{Authors withheld} 

% \author{Kuzmina Ekaterina, Razumov Artem, Oleg Rogov, Elfar Adalsteinsson, Jacob White, Dmitry V. Dylov} 
\author{Ekaterina Kuzmina\inst{1}
\and
Artem Razumov\inst{1}
\and
Oleg Y. Rogov\inst{1}
\and
Elfar Adalsteinsson\inst{2}
\and
Jacob White\inst{2}
\and
Dmitry V. Dylov\inst{1}
}
\authorrunning{E. Kuzmina et al.}

% First names are abbreviated in the running head.\orcidID{0000-0003-2251-3221}
% \institute{Affiliations withheld}
\institute{Skolkovo Institute of Science and Technology, Moscow, Russia\\
\and
Massachusetts Institute of Technology, Cambridge, MA, United States
}
\maketitle              % typeset the header of the contribution
\begin{abstract}
Image corruption by motion artifacts is an ingrained problem in Magnetic Resonance Imaging (MRI).
In this work, we propose a neural network-based regularization term to enhance \textit{Autofocusing}, a classic optimization-based method to remove motion artifacts.
The method takes the best of both worlds: the optimization-based routine iteratively executes the blind demotion and deep learning-based prior penalizes for unrealistic restorations and speeds up the convergence.
We validate the method on three models of motion trajectories, using synthetic and real noisy data.
The method proves resilient to noise and anatomic structure variation, outperforming the state-of-the-art demotion methods.

\end{abstract}
\section{Introduction}

% \subsection{Background}

Magnetic Resonance Imaging (MRI) non-invasively measures precise anatomical and functional patterns within the human body.
During the scan, it is nearly impossible for the patients to remain still in the scanner, with the movement of several millimeters or more being typical even for a healthy adult \cite{Friston1996}. 
Despite delivering superb detail and accelerating the acquisition,  powerful MRI machines are invariably prone to motion artifacts.
This stems from the voxels of the modern sensors, such as those in the high-resolution 7-Tesla scanners, so small that even a miniature patient movement results in severe voxel displacements \cite{havsteen2017movement}.

Generally, motion artifacts degrade the scan's quality, manifesting themselves as ghosting or blurring in the image \cite{Wood1985}.
The nature of these artifacts is convoluted, with many possible causing factors, ranging from the motion of an inner organ to that of the scanned body part, to the specifics of the \textit{k}-space trajectory in the signal acquisition protocol, to the patient's trembling, \textit{etc.} 
The periodic movements, such as cardiac motion and arterial pulsation, introduce distinctive well-defined patterns, whereas an arbitrary motion of the patient leads to a uniform image blur \cite{Zaitsev2015,2016motion}.

Motion artifacts lessen the overall diagnostic value of MRI data and may lead to misinterpretation when confused with pathology. The effect of motion alone, \textit{e.g.}, is estimated to introduce a 4\% error into the measurement of the gray matter in the brain  \cite{Reuter2015}.
That is especially crucial on the low signal-to-noise ratio (SNR) scans in the presence of noise or when the pathology is smaller than the voxel size of the scanner. 
Compensating for the motion artifacts is referred to as \textit{demotion} and is one of the most important tasks in the medical vision.

\noindent\subsubsection*{Related Work.}

The demotion methods can be broadly classified into \textit{prospective} (during the scan, \textit{e.g.}, \cite{Nehrke2005}) and \textit{retrospective} (after the scan, e.g., \cite{nielles2013vivo}).

In the latter category, ‘Autofocusing’ \cite{Atkinson1997} optimization algorithm became classic, allowing the motion parameters to be found \textit{iteratively} with respect to a chosen quality metric
and assuming the image corruption model is known.
Any such model, however, is hardly universal, given the omni-directional and non-uniform movements make the selection of a proper metric problematic \cite{McGee2000}.
Perception-based metrics could alleviate the challenge \cite{Tamada2020,Duffy2021}; however, MRI-specific standards for visual evaluation of motion artifacts do not exist \cite{Sommer2020}.

% CHANGED HERE!
Another powerful optimization-based Autofocusing method to remove motion artifacts is GradMC \cite{Loktyushin2015},
% Another powerful optimization-based  method to remove motion artifacts is GradMC \cite{Loktyushin2015},
the latest version of which compensates for \textit{nonrigid} physiological motions by distilling local \textit{rigid} motion of separate patches of the image. Instrumental in some cases, the method fails to handle sufficiently large movements and is computationally complex.
% CHANGED HERE!
In the presence of certain noise levels or when an object's structure varies abruptly GradMC underperforms and can introduce new artefacts.
% Similarly to Autofocusing, it underperforms in the presence of certain noise levels or when an object's structure varies abruptly.

With the advent of deep learning, the optimization methods were somewhat substituted by the convolutional neural networks (CNNs) \cite{Haskell2019,al2021stacked,kustner2019retrospective}, improving convergence of the non-convex motion parameter optimization procedure and/or increasing the reconstruction quality. A plethora of CNN-based \textit{deblurring} methods, such as popular DnCNN \cite{dncnn}, have then been implemented \cite{Sommer2020,Tamada2020}, without taking the physical nature of the MRI artifacts into account. Same applies to a generative adversarial network (GAN) approach MedGAN \cite{kustner2019retrospective}, entailing extra adversarial loss functions and the associated long training routine.

% Other approaches that is popular nowadays are based on neural networks, which are train to perform image deblurring.
% The article in \cite{Sommer2020} features a Foveal FCN architecture, the processing of the input image was performed at 3 different scales, which allowed to extract  high- and low-level features from the image. 
% In work \cite{Tamada2020} DCE-MRI images of liver were deblurred from respiratory motion corruption via neural network architecture based on the DnCNN architecture. % ref to out belowed Zhang
% MedGan architecture was used in the article \cite{kustner2019retrospective}, this approach include additional non-adversarial lossess and a modified generator architecture.
% In work \cite{oksuz2019detection} a CNN consisted of two blocks was used: a detector for defining data consistency term and a recurrent neural network (RNN) for image restoration. 
In the majority of works on MRI demotion, the corrupted data are either private \cite{kustner2019retrospective,Tamada2020} or generated by a model \cite{al2021stacked,Sommer2020,Duffy2021}. Herein, we also resort to the latter and employ a physics-based model to introduce the artifacts.

\begin{figure}[b] \label{fig:pipeline}
\includegraphics[width=\textwidth]{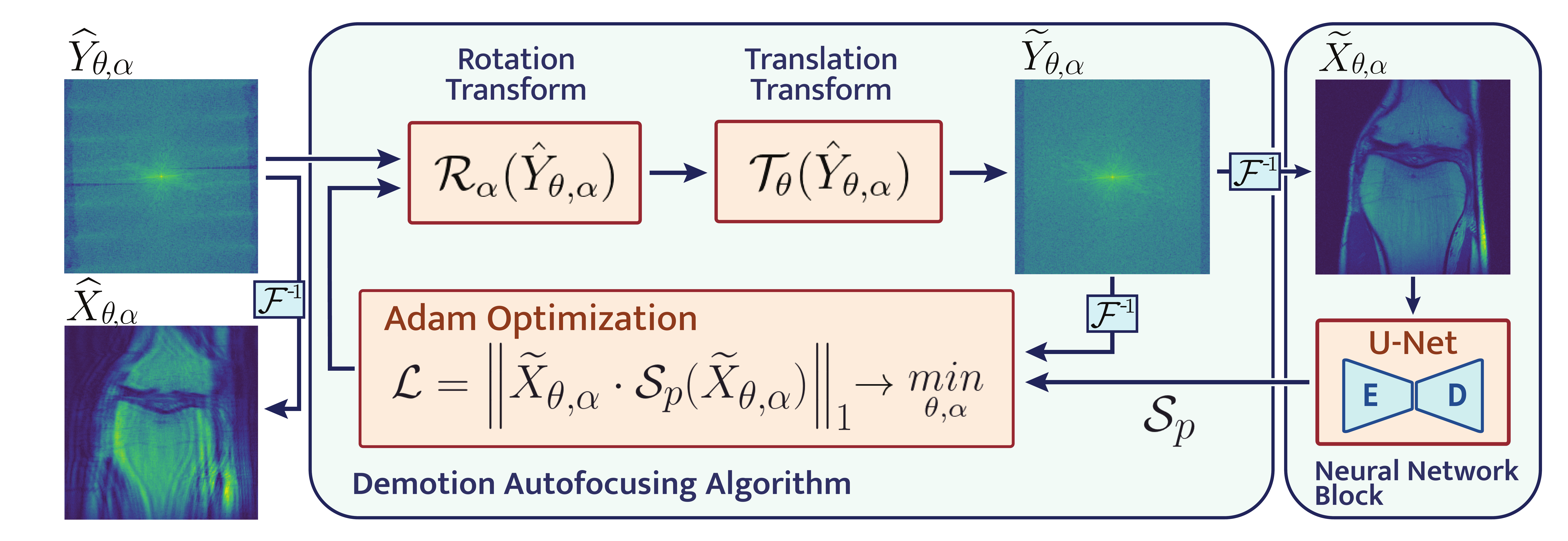}
\caption{Autofocusing+: proposed demotion method. 
The classic Autofocusing output $\widetilde{Y}_{\theta,\alpha}$ (left) is iteratively regularized with U-Net prior $\mathcal{S}_p(\widetilde{Y}_{\theta,\alpha})$(right). 
The L1-minimization of the regularized output updates the motion parameters $\theta$ and $\alpha$. 
% Then refined image $\widetilde{Y}_{\theta,\alpha}$ is passed to block with U-Net $\mathcal{S}_p$ and multiplied with it's output. 
% Finally, at the end of iteration step, L1-norm of the multiplication result $\widetilde{Y}_{\theta,\alpha} \cdot \mathcal{S}_p(\widetilde{Y}_{\theta,\alpha})$ is taken and motion parameters $\theta$ and $\alpha$ are updated.
}
\end{figure}

We propose to enhance the widely used Autofocusing algorithm by a CNN-extracted prior knowledge about \textit{k}-space specific motion corruption model, adding to the popular algorithm the new attributes of stability to anatomic structure variation and resilience to noise. Fig. \ref{fig:pipeline} summarizes the proposed method.

%%%%%%%%%%%
%%%%%%%%%%%
%%%%%%%%%%%
%%%%%%%%%%%
%%%%%%%%%%%
\section{Methods}\label{s:methods}
% \subsection{Model}\label{s:model}
\subsubsection{Motion Artifact Model.} 
\label{s:model}
% Unfortunately, nowadays it is hard to find a huge open-source dataset with real motion artifacts and spotted movement trajectories. 
% This requires large amount of data for training deep neural networks.
% Therefore, a model of generating synthetic data was developed for training of neural network. As it already has been mentioned, motion artifacts are an extremely complex problem and many aspects contribute to their formation. Furthermore, different types of motions result in different motion artifact appearance. 

\begin{figure}[t]
\includegraphics[width=\textwidth]{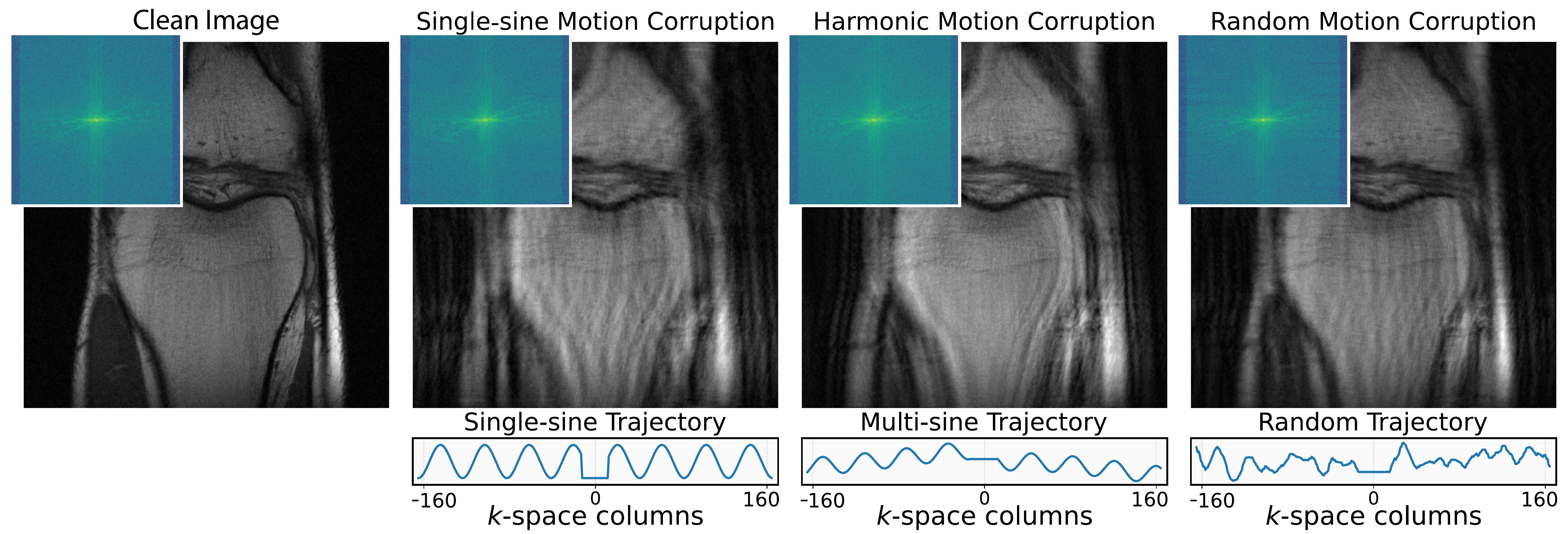}
\caption{MR image corrupted with different motion artifact trajectories in \textit{k}-space (insets show the corresponding spectra). Second raw: corresponding motion vectors. The movement trajectories considered: \textit{single-sine}, \textit{multi-sine} (or \textit{harmonic}), and \textit{random} (random contribution of multiple waves to the overall trajectory). 
} \label{fig:trajectory}
\end{figure}

We restrict the modeling to the problem of rigid body movements. 
The corrupted image $\hat{X}_{\theta, \alpha}$ can be obtained via sequential affine transforms of the clean image $X$: rotation $\mathcal{R}_{\alpha}(\cdot)$ and translation $\mathcal{T}_{\theta}(\cdot)$, characterized in the \textit{k}-space domain by the trajectory parameters $\alpha$ and $\theta$, respectively \cite{Atkinson1997}). Denoting $Y = \mathcal{F}(X)$ as the Fourier transform,

\begin{equation} \label{eq:corrupted_image} % ref: DONE
    \hat{X}_{\theta, \alpha} = \mathcal{F}^{-1}(\mathcal{R}_{\alpha}(\mathcal{T}_{\theta}(Y)))\,.
\end{equation}
% \subsubsection{Rotation motion.} \label{s:rotation}
% \textit{Fourier Rotation Theorem} implies that a rotation of a function $f(x, y)$ by an angle $\alpha$ implies that its Fourier transform $\mathcal{F}(u, v)$ is also rotated by the same angle $\alpha$ as $\mathcal{R}_{\alpha}\{f(x, y)\} \rightarrow \mathcal{R}_{\alpha}\{\mathcal{F}(u, v)\}$ \cite{larkin1997fast}.
% Given $ f(x, y) \supset \mathcal{F}\left(u, v\right)$, rotation can be performed as:
% \begin{multline}\label{e:rotation_theorem}
% f(x \cos \alpha + y \sin \alpha,-x \sin \alpha + y \cos \alpha) \supset \\ \mathcal{F}\left(u \cos \alpha+v \sin \alpha,-u \sin \alpha+v \cos \alpha\right)
% \end{multline}
% in \ref{e:rotation_theorem} (by angle from the motion trajectory $\alpha$) 

In practice, the \textit{k}-space rotation\footnote{Fourier Rotation Theorem \cite{larkin1997fast} allows for the equal rotation in the Image space.} is performed by multiplying each column of the coordinate grid $G$ by a rotation matrix:
% \ref{e:rotation_of_grid}:
% each column of coordinate grid $G$ was rotated by angle from the motion trajectory vector $\alpha$ via multiplication by a rotational matrix:

\begin{equation}\label{e:rotation_of_grid}
G_{\alpha}
=\left[\begin{array}{cc}
\cos \alpha & \sin \alpha \\
-\sin \alpha & \cos \alpha
\end{array}\right] \cdot 
G\,,
\end{equation}

\noindent which destroys the uniformity of the equispaced periodic frequencies and requires methods, such as Non-Uniform Fourier Transform (NUFFT) \cite{Fessler2003}, to place the frequency components on a new uniform grid correctly: % weiss2019pilot

\begin{equation}\label{eq:nufft}
\mathcal{R}_{\alpha}(Y) = \text{NUFFT}(Y, G_{\alpha})\,.
\end{equation}
% where $G_{\alpha}$ is a rotated by rotation \textit{k}-space sampling locations (grid) due to rotation $\alpha$
% \noindent NUFFT evaluates the coefficients of the Fourier series at specific locations. 
% In general it can be written as:
% \begin{equation}\label{e:nufft}
% Y_{j}=\mathcal{F}(X)_{j}=\sum_{k=-N / 2}^{N / 2-1} X_{k} e^{i t_{k} \cdot \omega_{j}}, \quad  j=-N / 2, \cdots, N / 2-1, N\in\mathbb{N}
% \end{equation}
% where 
%$\omega=\left\{\omega_{-N / 2}, \cdots, \omega_{N / 2-1}\right\}$ is a uniform frequency samples;  
% $\omega_{j}=2 \pi j / N \in[-\pi, \pi]$ is a non-uniform frequency samples and $t_{k}$ $\in[-N / 2, N / 2]$ is a uniform time samples.
%and $t=\left\{t_{-N / 2}, \cdots, t_{N / 2-1}\right\}$ not uniform time samples
% and time samples , $t_{k}$ $\in[-N / 2, N / 2]$. 
% \subsubsection{Translation Motion}\label{s:translation}

%  If $ X(x, y) \supset Y\left(f_{X}, f_{Y}\right)$, then $X(x-\theta) \supset \mathrm{e}^{-\mathrm{j} 2 \pi \theta s} Y(s)$ \cite{larkin1997fast}.
% \begin{equation}\label{e:translation}
%     X(x-\theta) \supset \mathrm{e}^{-\mathrm{j} 2 \pi \theta s} Y(s)  % here change exp power, add "where"
% \end{equation}

Likewise, the translation operation\footnote{Fourier Shift Theorem \cite{larkin1997fast} connects the pixel shifts to a phase change in \textit{k}-space.} can be formalized as:

\begin{equation}\label{e:translation}
    \mathcal{T}_{\theta}(Y) = |Y|\,\text{exp}(-j 2 \pi \angle Y + \theta)\,,
\end{equation}
where $\theta$ is the shift parameter,  $\angle Y$ and $|Y|$ are the angle and the magnitude of the complex values of $Y$.
The motion artifacts, generated by Eqns. (\ref{eq:nufft}) and (\ref{e:translation}), have to be reversible ($\mathcal{T}^{-1}_{\theta}(\mathcal{T}_{\theta}(X))=X$, $\mathcal{R}^{-1}_{\alpha}(\mathcal{R}_{\alpha}(X))=X$) and commutative ($\mathcal{T}_{\theta}(\mathcal{R}_{\alpha}(X))=\mathcal{R}_{\alpha}(\mathcal{T}_{\theta}(X))$). The motion vectors, set by various configurations of harmonic trajectories and parametrized by $\alpha$ and $\theta$, along with the modified \textit{k}-space and the resulting artifacts are shown in Fig.~\ref{fig:trajectory}.
%%%%%%%%
%%%%%%%%
%%%%%%%%
%%%%%%%%
\subsubsection{Autofocusing+ Algorithm.}\label{s:af}
% CHANGED HERE!
% The optimization task can be formulated as \textit{a blind demotion}, because the amplitude and the motion trajectory are unknown:
The optimization task of finding motion trajectories can be formulated as \textit{a blind demotion}, because the amplitude and the motion trajectory are unknown:
\begin{equation}\label{e:af_equation}
\argmin_{\theta, \alpha}\left \| \hat{X}_{\theta, \alpha} \right \|_{1}\,.
\end{equation}
% where $\theta$ and $\alpha$ are translation and rotation motion vectors parameters, respectively.
\noindent
% CHANGED HERE!
% For the optimization routine (ref. Algorithm \ref{alg:autofocusingplus}), Adam gradient descent \cite{Kingma2015} could be used 
For the optimization routine (ref. Algorithm \ref{alg:autofocusingplus}), previously conjugate gradient and LBFGS were used, but because of the computational complexity of our method, we apply basic Adam gradient descent \cite{Kingma2015}
to iteratively update the motion vectors \textit{w.r.t.} the image quality metric. 
Inspired by previous studies \cite{lustig2008compressed}\cite{Yang2013}, L1-norm was used as the image quality measure $\mathcal{L}_{AF}$. 
The L1-norm choice is rational because the motion artifacts blur sharp edges in the image and, therefore, decrease the image sparsity by `redistributing' intensity values between the neighboring pixels\footnote{Our experimentation confirmed that L1-norm is a superior choice, outperforming the other metrics
used with Autofocusing algorithms \cite{McGee2000} (not shown).}.
Below, the Adam-based autofocusing algorithm is referred to as `baseline Autofocusing'. 

% Algorithmic BEGIN
\begin{algorithm}[b]
  \caption{Autofocusing+ Optimization}
  \label{alg:autofocusingplus}
\begin{algorithmic}
    \INPUT{$\hat{Y}_{\theta, \alpha} \in \mathbb{C}^{n \times n}$ // \textit{k}-space corrupted by motion with parameters 
    $\theta, \alpha $}
    \OUTPUT{$\tilde{Y}_{\theta, \alpha} \in \mathbb{C}^{n \times n}$ // Refined \textit{k}-space}
    \FOR {train epoch}
    \FOR {optimization step}
    \STATE {Compute image quality metric} $\mathcal{L}_{AF} = \left \| \hat{X}_{\theta, \alpha} \cdot \mathcal{S}_p(\hat{X}_{\theta, \alpha}) \right \|_{1}$
    
    \STATE {Calculate gradients for Autofocusing parameters $\theta$, $\alpha$}
    \STATE {Update Autofocusing parameters $\theta, \alpha$ with \textit{Adam Gradient Descent}} 

    \ENDFOR
    \STATE {Compute loss between target and refined image} $\mathcal{L}_{NN}(\hat{X}_{\theta, \alpha}, X) $
    \STATE {Calculate gradients for Neural Network parameters $p$}
    \STATE {Update Neural Network parameters $p$ with \textit{Adam Gradient Descent}} 
    \ENDFOR
\end{algorithmic}
% \vskip -0.3in
\end{algorithm}
% Algorithmic END

\subsubsection{U-Net as an Additional Prior.} 
Our idea is to use a CNN alongside the image quality metric in the optimization procedure to improve the image restoration quality via an additional image prior \cite{kingma2013auto,siddique2021u}. Naturally, a simple U-Net \cite{Ronneberger} architecture was considered for such a regularization task. 

Importantly, in the proposed pipeline (See Fig.\ref{fig:pipeline}), the CNN is the part of the minimized function, the parameters of which are also trainable. That is, we propose to use CNN as a part of loss and then learn the parameters of this loss in the Autofocusing algorithm.
As such, given the refined image $\hat{X}_{\theta, \alpha}$ and the output of the network $\mathcal{S}_p(\hat{X}_{\theta, \alpha})$, the optimization problem can be re-written as

\begin{equation}\label{eq:argmin}
\theta(p), \alpha(p) = \argmin_{\theta, \alpha}\left \| \hat{X}_{\theta, \alpha} \cdot \mathcal{S}_p(\hat{X}_{\theta, \alpha}) \right \|_{1}\,,
\end{equation}
where $\mathcal{S}_p(\cdot)$ and $p$ are the CNN inference output and its weights, respectively.
For consistency between various extents of corruption by the motion, a
sigmoid is applied to the final layer of the network to scale all its values from 0 to 1. Thus, the output of the network scales the corrupted image prior to the L1-norm.

\begin{figure}[h]
\label{fig:exampl_results}
\includegraphics[width=\textwidth]{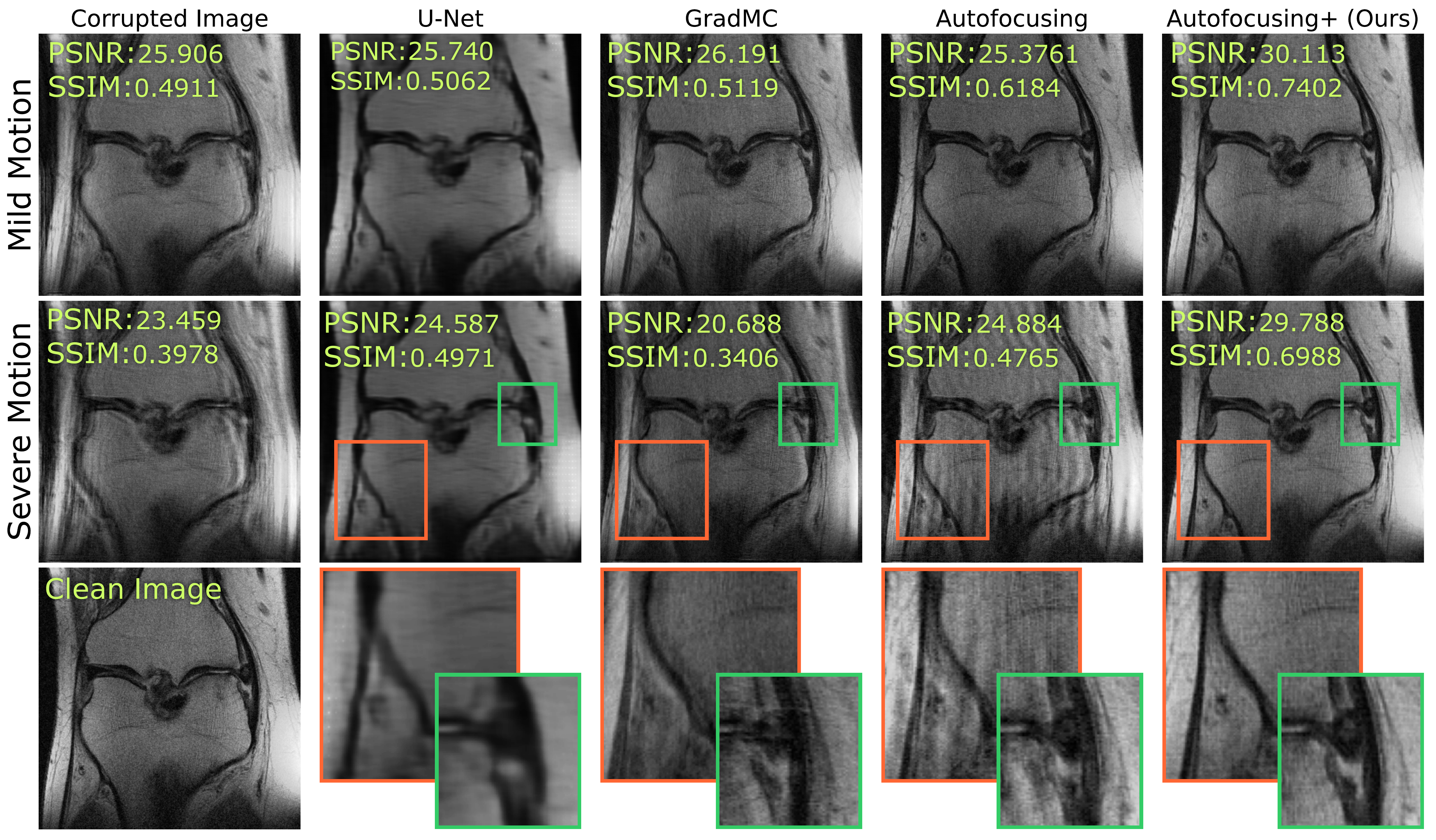}
\caption{Comparison of different demotion methods for mild (first row) and severe (second row) corruption with harmonic trajectories for the data from low-SNR MRI machine (1T, real noise). The brightness was adjusted for viewer's convenience.} 
\end{figure}

% \subsubsection{Neural Network Block.}\label{s:train}
% Since we aim to demonstrate how neural network can improve Autofocusing optimization procedure, we do not compare different segmentation architectures inside our demotion framework \ref{fig:pipeline} and used 
Basic U-Net architecture with the blocks consisting of Instance Norm and Leaky ReLu  is then trained as follows: % \cite{xu2015empirical}

\begin{equation}\label{eq:unet_train}
p = \argmin_p \left \| X -  \mathcal{R}^{-1}_{\alpha(p)}\left(\mathcal{T}^{-1}_{\theta(p)}(\hat{X}_{\theta, \alpha})\right)\right \|_{1}\,,
\end{equation}
introducing new hyperparameters: the number of autofocusing steps in Algorithm \ref{alg:autofocusingplus} and the learning rate. Figs. \ref{fig:exampl_results}, \ref{fig:box_plot_mild}, \ref{fig:noise_results} showcase the algorithm's performance.
%
%%%%%
%%%%%
%%%%%
%%%%%
\section{Experiments}\label{s:experiments}
\textbf{Data.}\label{s:data}
\texttt{fastMRI} dataset with the raw \textit{k}-space data was used \cite{zbontar2018fastmri}. 
To train and evaluate the model, the single-coil proton density weighted knee scans, acquired at 3T and 1T, were selected, including 10,374 scans for training and 1,992 scans for validation.
% It contains original \textit{k}-space of brains and knees in different modalities.
% Datasets consist of PD-weighted (proton density) images only. Such images has the least noise, which helps to keep the entire dataset consistent. 
Images were cropped to the size of 320$\times$320 pixels and then corrupted with the motion artifact model described in Sec. \ref{s:model}.

\smallskip
\noindent\textbf{Metrics.}\label{s:metrics}
The reconstructed images are compared using four metrics: PSNR, SSIM \cite{Wang2004}, MS-SSIM \cite{wang2003multiscale}, and VIF (Visual Information Fidelity \cite{Sheikh2006}).
% Visual Information Fidelity (VIF)\cite{Sheikh2006}
% VIF is a quality assessment index based on natural scene statistics and the notion of image information extracted by the human visual system.
% Multiscale Structural structural similarity index (MS SSIM) \cite{wang2003multiscale}
% MS SSIM is an extension of SSIM. It supplies more flexibility than previous single-scale methods in incorporating the variations of viewing conditions.

\noindent \textbf{Implementation Details.} To train U-Net inside the algorithm, Adam optimizer was used with $\beta_1$ = 0.900, $\beta_2$ = 0.999, and the learning rate $lr$ = 5$\times 10^{-5}$.
The gradients are accumulated at each iteration of Autofocusing algorithm, with the total number of steps set to 30, and with the rate of updating the motion parameters $lr$ = 3$\times 10^{-4}$.
When the Autofocusing optimization ends, L1-loss between the ground truth and the refined image is taken and the error is backpropagated to optimize the weights of U-Net. 
Thus, the dependence on parameters $\theta, \alpha$, and the weights $p$, implicit in Eqns. (\ref{eq:unet_train}) and (\ref{eq:argmin}), is examined. 
On average, it takes 160 epochs to converge.
No augmentation was required.

The upper bound for the motion vector was set to 2 degree rotations and 5 pixel translations, and the corruption model dismissed the central 8\% of the \textit{k}-space, ensuring the low frequencies, responsible for the content, are preserved \cite{Loktyushin2015}. 
Random motion vector was generated from the Gaussian distribution and then smoothed with Savitzky–Golay filter \cite{Savitzky1964} with the kernel size of 20.

% The size of training dataset contained 50 images with milti-sine (harmonic) motion corruption. 
% We ran extensive series of experiments outlined in Table \ref{s:all_metrics}. Experimental results are provided in the format of $\mu \pm \sigma$, where $\mu$ is the average metric value over the hold-out set, and $\sigma$ is its standard deviation.
% Experiments confirmed that our model satisfies the criteria described in Sections \ref{s:rotation} and \ref{s:translation}  with a computational error less than 1\% for all quality metrics.
% Ablation study is provided in the Supplement.

\section{Results}\label{s:results}

\begin{figure}[h] \label{fig:box_plot_mild}
\includegraphics[width=\textwidth]{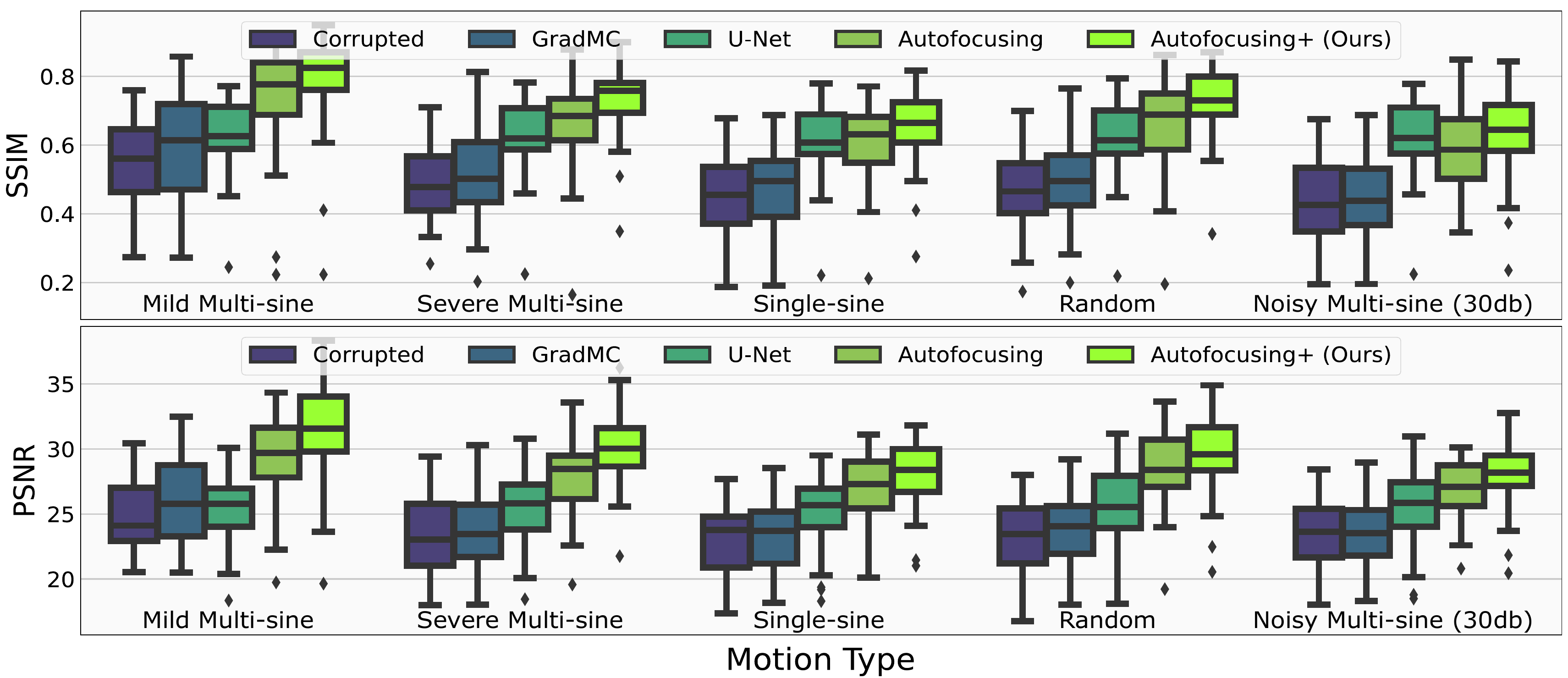}
\caption{Mild corruption demotion via different methods. 3T scans and synthetic noise.
Autofocusing+ differs from Autofocusing for all data significantly (max p-value \textless 10$^{-5}$).
} 
\end{figure}

The comparison between the state-of-the-art (SOTA) optimization- and deep learning-based algorithms is shown in Fig. \ref{fig:exampl_results} for mild and severe motion corruptions.
The validation of our method was performed with all three types of motion trajectories shown in Fig. \ref{fig:trajectory}. A full comparison of the trajectories is presented in Fig. \ref{fig:box_plot_mild}, suggesting that the technique is functional for each of them, even the random one. 
We refer to Supplementary material for all four quality metrics and each motion trajectory (Table \ref{s:all_metrics}), while presenting only selected metrics herein for brevity, with the focus on the harmonic corruption as the most realistic \cite{kustner2019retrospective}.

\begin{figure}[t]
\includegraphics[width=\textwidth]{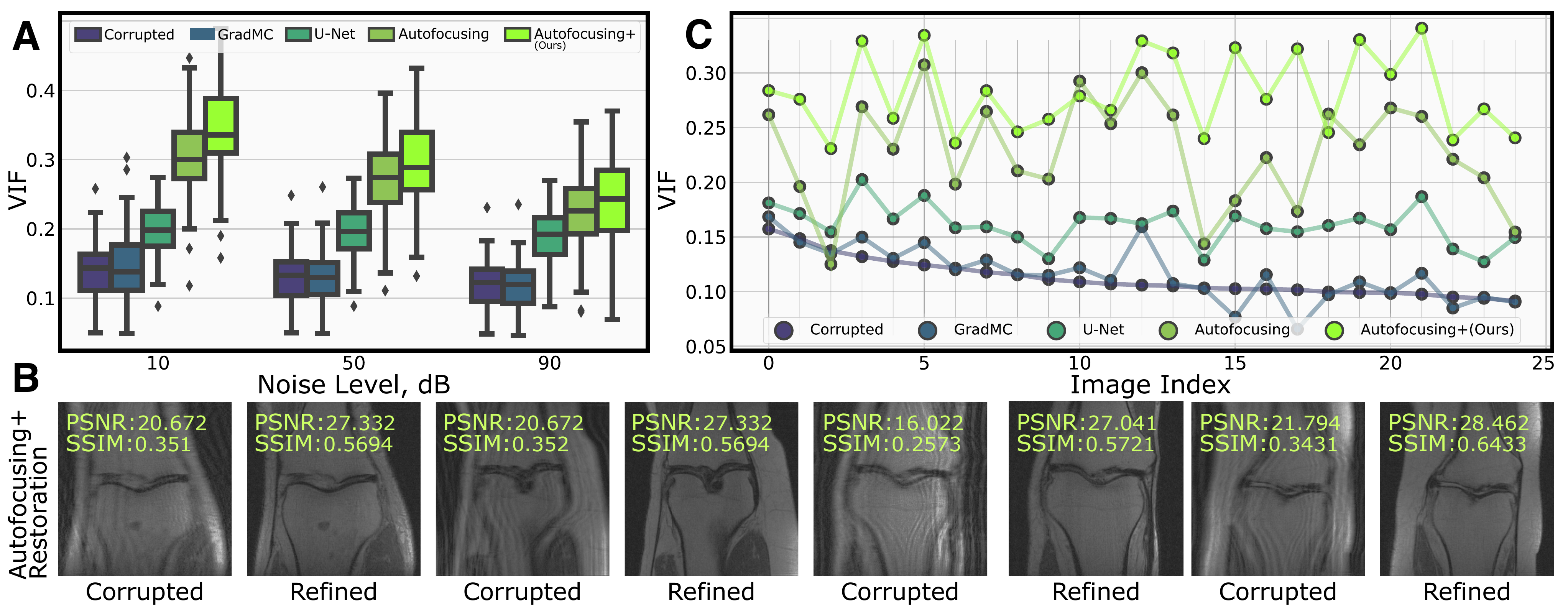}
\caption{Demotion results for harmonic motion corruption and real noise (1T scans). \textbf{A.} Performance for different noise levels (max p-value \textless 0.0047). \textbf{B.} Example of real noisy images restored by Autofocusing+. \textbf{C.} Performance of all methods on 25 random images in the real-noise dataset. 
In some cases, classic Autofocusing fails to restore the image (\textit{e.g.}, image \#2, \#14, and \#17), whereas Autofocusing+ handles the demotion.} 
\label{fig:noise_results}
\end{figure}

\smallskip
\noindent\textbf{Resilience to Noise.} % mcrobbie2007mri
The demotion methods have to be resilient to noise (\textit{e.g.}, electronic \cite{brown2014magnetic}) and abrupt fluctuations in the detail of an anatomic structure (\textit{e.g.}, when the ligaments of the joint appear next to the motion artifact). These cases are thoroughly studied in Fig. \ref{fig:noise_results}, where either classic optimization or deep learning methods sometimes fail to function.
On the opposite, Autofocusing+ is consistent and robust on all images, showing resilience to the strongest noise levels (90 dB).
In the synthetic noise experiments, the noise is modeled as Gaussian distribution $\mathcal{N}(\mu,\sigma^{2})$ added to the signal in \textit{k}-space: $Y' = Y + \mathcal{N}(\mu,\sigma^{2})$ \cite{mohan2014survey}, where $Y'$ is a noise-corrupted spectrum, $\mu$ is the mean, and $\sigma$ is the standard deviation of the noise.

We also validated our algorithm on a real-noise data subset, a part of \texttt{fastMRI} acquired at 1T, where the SNR and the contrast are visually sub-optimal (the factual noise level is unknown). Although all methods fluctuate more severely than in the case of synthetic noise, the t-test confirms that our algorithm still outperforms the second best method (classic Autofocusing) with statistical significance (max p-value \textless 0.0047).

%
% To approve the difference between Autofocusing and Autofocusing+ algorithms statistically for Fig. \ref{fig:noise_results}, A., one sample t-test on dependent pairs of image metrics differences was conducted for each level of noise. We discovered a positive and significant improvement in metrics for all 3 levels of noise presented in fig. \ref{fig:noise_results} (maximal p-value \textless 0.0047)

% \vspace{-1cm}
\section{Discussion}\label{s:discuss} 
As can be seen in Figs. \ref{fig:box_plot_mild} and \ref{fig:noise_results}, Autofocusing+ always outperforms the SOTA models for all four metrics, all noise levels, and all motion trajectories. This affirms that the addition of neural network helps to regularize the optimization problem and improves the performance of the baseline Autofocusing method.
Even visually (Fig.\ref{fig:noise_results}(C)), Autofocusing+ succeeds in the cases where the classic optimization alone fails to refine the detail and even introduces new artifacts.
We observe that such new artifacts correspond to abrupt variation of pixel intensity in the anatomic detail, where the structure `couples' with the motion artifacts. 
The use of the neural network, however, extracts \textit{prior knowledge} for the image quality measure and eliminates the issue, making the reconstruction stable and the image well-refined. 

\begin{figure} 
\includegraphics[width=\textwidth]{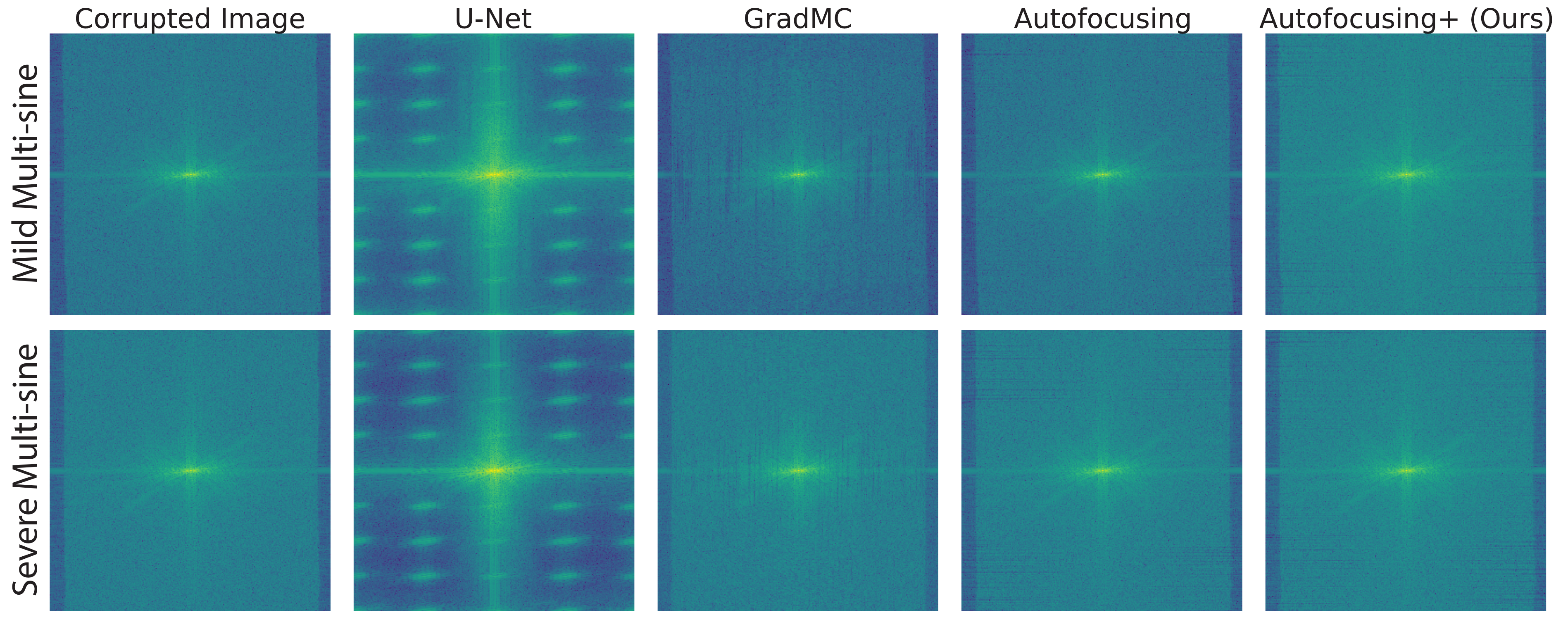}
\caption{\textit{k}-space of refined images.} \label{fig:restored_kspace} % 
\end{figure}

The evaluation of the reconstructed knee images for \textit{mild} motion harmonic trajectories in Fig. \ref{fig:exampl_results} and Fig. \ref{fig:restored_kspace} reveals only subtle differences in the performance of various methods.
However, the \textit{severe} corruption makes the classical optimization-based approaches inapt for removing the artifacts, whereas Autofocusing+ refines image detail, making the fine structure acceptable visually and leaving the uniform regions intact.
According to Fig.\ref{fig:noise_results}, Autofocusing+ is also efficient in the demotion task in the presence of noise, being functional both for high levels of added noise and for real low-SNR noisy MRI scans. 

It can also be noticed that a simple deblurring with a stand-alone U-Net yields approximately the same consistent output for different noise levels and the strengths of motion corruption (Table \ref{s:all_metrics} in Supplementary material), learning to `smooth out' the motion artefact patterns instead of fixing them \cite{siddique2021u,pronina2020microscopy}. This results in an irreversible loss of the fine tissue structure, making the use of optimization-based demotion methods more justified and preferred in practice.

Hence, our method takes the best of both worlds: the optimization-based routine iteratively performs the blind demotion with interpretable measure of the image detail and the deep learning-based prior helps to penalize for unrealistic restorations. The latter is especially useful in practice as it helps prevent the emergence of new artifacts, sometimes induced by neural networks in MRI reconstructions \cite{belov2021towards}.

%%%%%
%%%%%
%%%%%
\smallskip
\textit{Limitations.} 
Similarly to other synthetic motion works \cite{Duffy2021,Loktyushin2015,oksuz2018deep}, a Cartesian sampling with 2D in-plane motions in the phase encoding direction was assumed. 
Elastic movements (\textit{e.g.}, compression and stretching in the soft tissues \cite{2016motion}) were beyond the scope of our work and should be revisited, following, \textit{e.g.}, \cite{Zaitsev2015}.

Stemming from the baseline Autofocusing, and similarly to other optimization-based methods, there is room to improve \textit{the inference time} of Autofocusing+ which takes 7--9 minutes to process a batch of 10 images. Parallel computing and the pertinent efficient acceleration of the gradients and the Fourier transforms \cite{shipitsin2020global} should work well and will be the subject of future work. Likewise, the study of different neural network architectures should be conducted.
\vspace{-0.5cm}
\subsubsection{Conclusion.}
In this work, we presented Autofocusing+, a new deep learning-based regularization method to enhance the classic optimization algorithm for compensating the corruption caused by the subject's motion.
The blind demotion routine uses the regularizer as an image prior penalizing for unrealistic restorations and improving the reconstruction quality.
We validated the method on three models of motion trajectories, using synthetic and real noisy data, showing resilience to noise and anatomic structure variation and outperforming the SOTA demotion methods.

% \section{Acknowledgment}
% Withheld.

\bibliographystyle{splncs04}
\bibliography{biblio}

\begin{thebibliography}{10}
\providecommand{\url}[1]{\texttt{#1}}
\providecommand{\urlprefix}{URL }
\providecommand{\doi}[1]{https://doi.org/#1}

\bibitem{al2021stacked}
Al-masni, M.A., Lee, S., Yi, J., Kim, S., Gho, S.M., Choi, Y.H., Kim, D.H.:
  Stacked u-nets with self-assisted priors towards robust correction of rigid
  motion artifact in brain mri. arXiv preprint arXiv:2111.06401  (2021)

\bibitem{Atkinson1997}
Atkinson, D., Hill, D.L., Stoyle, P.N., Summers, P.E., Keevil, S.F.: Automatic
  correction of motion artifacts in magnetic resonance images using an entropy
  focus criterion. IEEE Transactions on Medical Imaging  \textbf{16} (1997)

\bibitem{belov2021towards}
Belov, A., Stadelmann, J., Kastryulin, S., Dylov, D.V.: Towards ultrafast mri
  via extreme k-space undersampling and superresolution. In: International
  Conference on Medical Image Computing and Computer-Assisted Intervention. pp.
  254--264. Springer, Cham (2021)

\bibitem{brown2014magnetic}
Brown, R.W., Cheng, Y.C.N., Haacke, E.M., Thompson, M.R., Venkatesan, R.:
  Magnetic resonance imaging: physical principles and sequence design. John
  Wiley \& Sons (2014)

\bibitem{Duffy2021}
Duffy, B.A., Zhao, L., Sepehrband, F., Min, J., Wang, D.J., Shi, Y., Toga,
  A.W., Kim, H.: Retrospective motion artifact correction of structural mri
  images using deep learning improves the quality of cortical surface
  reconstructions. NeuroImage  \textbf{230} (2021)

\bibitem{Fessler2003}
Fessler, J.A., Sutton, B.P.: Nonuniform fast fourier transforms using min-max
  interpolation. IEEE Transactions on Signal Processing  \textbf{51} (2003)

\bibitem{Friston1996}
Friston, K.J., Williams, S., Howard, R., Frackowiak, R.S., Turner, R.:
  Movement-related effects in fmri time-series. Magnetic Resonance in Medicine
  \textbf{35} (1996)

\bibitem{2016motion}
Godenschweger, F., K{\"a}gebein, U., Stucht, D., Yarach, U., Sciarra, A.,
  Yakupov, R., L{\"u}sebrink, F., Schulze, P., Speck, O.: Motion correction in
  mri of the brain. Physics in medicine \& biology  \textbf{61}(5), ~R32 (2016)

\bibitem{Haskell2019}
Haskell, M.W., Cauley, S.F., Bilgic, B., Hossbach, J., Splitthoff, D.N.,
  Pfeuffer, J., Setsompop, K., Wald, L.L.: Network accelerated motion
  estimation and reduction (namer): Convolutional neural network guided
  retrospective motion correction using a separable motion model. Magnetic
  Resonance in Medicine  \textbf{82} (2019)

\bibitem{havsteen2017movement}
Havsteen, I., Ohlhues, A., Madsen, K.H., Nybing, J.D., Christensen, H.,
  Christensen, A.: Are movement artifacts in magnetic resonance imaging a real
  problem?—a narrative review. Frontiers in neurology  \textbf{8}, ~232
  (2017)

\bibitem{Kingma2015}
Kingma, D.P., Ba, J.L.: Adam: A method for stochastic optimization (2015)

\bibitem{kingma2013auto}
Kingma, D.P., Welling, M.: Auto-encoding variational bayes. arXiv preprint
  arXiv:1312.6114  (2013)

\bibitem{kustner2019retrospective}
K{\"u}stner, T., Armanious, K., Yang, J., Yang, B., Schick, F., Gatidis, S.:
  Retrospective correction of motion-affected mr images using deep learning
  frameworks. Magnetic resonance in medicine  \textbf{82}(4),  1527--1540
  (2019)

\bibitem{larkin1997fast}
Larkin, K.G., Oldfield, M.A., Klemm, H.: Fast fourier method for the accurate
  rotation of sampled images. Optics communications  \textbf{139}(1-3),
  99--106 (1997)

\bibitem{Loktyushin2015}
Loktyushin, A., Nickisch, H., Pohmann, R., Schölkopf, B.: Blind multirigid
  retrospective motion correction of mr images. Magnetic Resonance in Medicine
  \textbf{73} (2015)

\bibitem{lustig2008compressed}
Lustig, M., Donoho, D.L., Santos, J.M., Pauly, J.M.: Compressed sensing mri.
  IEEE signal processing magazine  \textbf{25}(2),  72--82 (2008)

\bibitem{McGee2000}
McGee, K.P., Manduca, A., Felmlee, J.P., Riederer, S.J., Ehman, R.L.: Image
  metric-based correction (autocorrection) of motion effects: Analysis of image
  metrics. Journal of Magnetic Resonance Imaging  \textbf{11} (2000)

\bibitem{mohan2014survey}
Mohan, J., Krishnaveni, V., Guo, Y.: A survey on the magnetic resonance image
  denoising methods. Biomedical signal processing and control  \textbf{9},
  56--69 (2014)

\bibitem{Nehrke2005}
Nehrke, K., Börnert, P.: Prospective correction of affine motion for arbitrary
  mr sequences on a clinical scanner. Magnetic Resonance in Medicine
  \textbf{54} (2005)

\bibitem{nielles2013vivo}
Nielles-Vallespin, S., et~al.: In vivo diffusion tensor mri of the human heart:
  reproducibility of breath-hold and navigator-based approaches. Magnetic
  resonance in medicine  \textbf{70}(2),  454--465 (2013)

\bibitem{oksuz2018deep}
Oksuz, I., et~al.: Deep learning using k-space based data augmentation for
  automated cardiac mr motion artefact detection. In: International Conference
  on Medical Image Computing and Computer-Assisted Intervention. pp. 250--258.
  Springer (2018)

\bibitem{pronina2020microscopy}
Pronina, V., Kokkinos, F., Dylov, D.V., Lefkimmiatis, S.: Microscopy image
  restoration with deep wiener-kolmogorov filters. In: European Conference on
  Computer Vision. pp. 185--201. Springer, Cham (2020)

\bibitem{Reuter2015}
Reuter, M., et~al.: Head motion during mri acquisition reduces gray matter
  volume and thickness estimates. NeuroImage  \textbf{107} (2015)

\bibitem{Ronneberger}
Ronneberger, O., et~al.: U-net: Convolutional networks for biomedical image
  segmentation. In: MICCAI 2015. pp. 234--241. Springer International
  Publishing, Cham (2015)

\bibitem{Savitzky1964}
Savitzky, A., Golay, M.J.: Smoothing and differentiation of data by simplified
  least squares procedures. Analytical Chemistry  \textbf{36} (1964)

\bibitem{Sheikh2006}
Sheikh, H.R., Bovik, A.C.: Image information and visual quality. IEEE
  Transactions on Image Processing  \textbf{15} (2006)

\bibitem{shipitsin2020global}
Shipitsin, V., Bespalov, I., Dylov, D.V.: Global adaptive filtering layer for
  computer vision. arXiv preprint arXiv:2010.01177  (2020)

\bibitem{siddique2021u}
Siddique, N., Paheding, S., Elkin, C.P., Devabhaktuni, V.: U-net and its
  variants for medical image segmentation: A review of theory and applications.
  IEEE Access  (2021)

\bibitem{Sommer2020}
Sommer, K., et~al.: Correction of motion artifacts using a multiscale fully
  convolutional neural network. American Journal of Neuroradiology  \textbf{41}
  (2020)

\bibitem{Tamada2020}
Tamada, D., Kromrey, M.L., Ichikawa, S., Onishi, H., Motosugi, U.: Motion
  artifact reduction using a convolutional neural network for dynamic contrast
  enhanced mr imaging of the liver. Magnetic Resonance in Medical Sciences
  \textbf{19} (2020)

\bibitem{Wang2004}
Wang, Z., Bovik, A.C., Sheikh, H.R., Simoncelli, E.P.: Image quality
  assessment: From error visibility to structural similarity. IEEE Transactions
  on Image Processing  \textbf{13} (2004)

\bibitem{wang2003multiscale}
Wang, Z., Simoncelli, E.P., Bovik, A.C.: Multiscale structural similarity for
  image quality assessment. In: The Thrity-Seventh Asilomar Conference on
  Signals, Systems \& Computers, 2003. vol.~2, pp. 1398--1402. Ieee (2003)

\bibitem{Wood1985}
Wood, M.L., Henkelman, R.M.: Mr image artifacts from periodic motion. Medical
  Physics  \textbf{12} (1985)

\bibitem{Yang2013}
Yang, Z., Zhang, C., Xie, L.: Sparse mri for motion correction (2013)

\bibitem{Zaitsev2015}
Zaitsev, M., Maclaren, J., Herbst, M.: Motion artifacts in mri: A complex
  problem with many partial solutions (2015)

\bibitem{zbontar2018fastmri}
Zbontar, J., , et~al.: fastmri: An open dataset and benchmarks for accelerated
  mri. arXiv preprint arXiv:1811.08839  (2018)

\bibitem{dncnn}
Zhang, K., Zuo, W., Chen, Y., Meng, D., Zhang, L.: Beyond a gaussian denoiser:
  Residual learning of deep cnn for image denoising. IEEE Transactions on Image
  Processing  \textbf{26}(7),  3142--3155 (2017)

\end{thebibliography}

\clearpage
\newpage
\section{Supplementary}

\renewcommand{\thetable}{S\arabic{table}}

\setcounter{table}{0}    

\begin{table}[h!] 
\fontsize{8}{8}\selectfont
\centering
\caption{Demotion results on validation datasets for various motion trajectories, noise inputs and methods.}
\begin{tabular}{c c@{\hskip 0.1in} c@{\hskip 0.1in} c@{\hskip 0.1in} c@{\hskip 0.1in} c} 
 \toprule
Motion Type &  Method & PSNR & SSIM & VIF & MS-SSIM \\ [0.5ex] 
 \midrule
 Harmonic Mild & Corrupted & 24.98 $\pm$ 2.60 & 0.554 $\pm$ 0.11 & 0.186 $\pm$ 0.06 & 0.873 $\pm$ 0.05\\
 &GradMC & 26.06 $\pm$ 3.23 & 0.595 $\pm$ 0.15 & 0.216 $\pm$ 0.09 & 0.895 $\pm$ 0.06\\ 
 &U-Net & 25.51 $\pm$ 2.48 & 0.629 $\pm$ 0.09 & 0.196 $\pm$ 0.04 & 0.890 $\pm$ 0.03 \\
 &Autofocusing & 29.46 $\pm$ 3.02 & 0.740 $\pm$ 0.14 & 0.373 $\pm$ 0.09 & 0.963 $\pm$ 0.03 \\
& \textbf{Autofocusing+(Ours)} & \textbf{31.46 $\pm$ 3.36} & \textbf{0.792 $\pm$ 0.13} & \textbf{0.407 $\pm$ 0.10} & \textbf{0.972 $\pm$ 0.03} \\ [1ex] 
 \midrule
 Harmonic Severe & Corrupted & 23.14 $\pm$ 2.75 & 0.486 $\pm$ 0.10 & 0.142 $\pm$ 0.04 & 0.804 $\pm$ 0.06\\
 &GradMC & 23.71 $\pm$ 2.81 & 0.512 $\pm$ 0.12 & 0.148 $\pm$ 0.06 & 0.819 $\pm$ 0.07\\ 
 &U-Net & 25.57 $\pm$ 2.68 & 0.626 $\pm$ 0.09 & 0.195 $\pm$ 0.04 & 0.889 $\pm$ 0.03 \\
 &Autofocusing & 27.91 $\pm$ 2.55 & 0.664 $\pm$ 0.12 & 0.305 $\pm$ 0.06 & 0.946 $\pm$ 0.03 \\
 &\textbf{Autofocusing+(Ours)} & \textbf{30.08 $\pm$ 2.56} & \textbf{0.735 $\pm$ 0.09} & \textbf{0.349 $\pm$ 0.06} & \textbf{0.964 $\pm$ 0.02} \\ [1ex] 
  \midrule
Single-sine & Corrupted & 22.90 $\pm$ 2.57 & 0.458 $\pm$ 0.10 & 0.132 $\pm$ 0.03 & 0.780 $\pm$ 0.05\\
 & GradMC & 23.33 $\pm$ 2.48 & 0.482 $\pm$ 0.10 & 0.136 $\pm$ 0.04 & 0.793 $\pm$ 0.06\\ 
 & U-Net & 25.27 $\pm$ 2.65 & 0.613 $\pm$ 0.09 & 0.183 $\pm$ 0.03 & 0.879 $\pm$ 0.03 \\
 & Autofocusing & 27.27 $\pm$ 2.30 & 0.608 $\pm$ 0.10 & 0.261 $\pm$ 0.04 & 0.924 $\pm$ 0.03 \\ 
 & \textbf{Autofocusing+(Ours)} & \textbf{28.15 $\pm$ 2.48} & \textbf{0.653 $\pm$ 0.10} & \textbf{0.285 $\pm$ 0.05} & \textbf{0.939 $\pm$ 0.03} \\ [1ex] 
 \midrule
 Random & Corrupted & 23.26 $\pm$ 2.68 & 0.468 $\pm$ 0.10 & 0.136 $\pm$ 0.04 & 0.786 $\pm$ 0.06\\  
  & GradMC & 23.89 $\pm$ 2.75 & 0.501 $\pm$ 0.11 & 0.145 $\pm$ 0.05 & 0.810 $\pm$ 0.07\\  
 & U-Net & 25.50 $\pm$ 2.74 & 0.621 $\pm$ 0.09 & 0.189 $\pm$ 0.04 & 0.884 $\pm$ 0.03 \\ 
 & Autofocusing & 28.68 $\pm$ 2.75 & 0.665 $\pm$ 0.13 & 0.303 $\pm$ 0.07 & 0.942 $\pm$ 0.04 \\
 & \textbf{Autofocusing+(Ours)} & \textbf{29.83 $\pm$ 2.94} & \textbf{0.723 $\pm$ 0.11} & \textbf{0.341 $\pm$ 0.07} & \textbf{0.958 $\pm$ 0.03} \\ [1ex] 
  \midrule
 \makecell{Harmonic with \\ noise 30 dB} & Corrupted & 23.35 $\pm$ 2.66 & 0.440 $\pm$ 0.11 & 0.136 $\pm$ 0.04 & 0.800 $\pm$ 0.06\\
  & GradMC & 23.46 $\pm$ 2.61 & 0.449 $\pm$ 0.11  & 0.135 $\pm$ 0.04 & 0.802 $\pm$ 0.06\\ 
 & U-Net & 25.67 $\pm$ 2.70 & 0.625 $\pm$ 0.09 & 0.195 $\pm$ 0.04 & 0.889 $\pm$ 0.03 \\
 & Autofocusing & 26.86 $\pm$ 2.13 & 0.586 $\pm$ 0.11 & 0.270 $\pm$ 0.06 & 0.927 $\pm$ 0.03 \\
 & \textbf{Autofocusing+(Ours)} & \textbf{28.19 $\pm$ 2.49} & \textbf{0.640 $\pm$ 0.12} & \textbf{0.300 $\pm$ 0.06} & \textbf{0.942 $\pm$ 0.03} \\  [1ex]
   \midrule
 Real noisy with & Corrupted & 22.81 $\pm$ 2.23 & 0.379 $\pm$ 0.09 & 0.810 $\pm$ 0.05 & 0.131 $\pm$ 0.03\\
mild harmonic & GradMC & 23.28  $\pm$  2.26 & 0.399 $\pm$  0.13 & 0.140  $\pm$  0.05 & 0.814  $\pm$  0.07\\ 
 &U-Net & 23.92  $\pm$  2.39 & 0.442  $\pm$  0.08 & 0.148  $\pm$  0.02 & 0.840  $\pm$  0.04 \\
 &Autofocusing & 26.03  $\pm$  2.77 & 0.590  $\pm$  0.15 & 0.271  $\pm$  0.08 & 0.931  $\pm$  0.04 \\
 &\textbf{Autofocusing+(Ours)} & \textbf{26.97  $\pm$  2.26} & \textbf{0.631  $\pm$  0.07} & \textbf{0.289  $\pm$  0.04} & \textbf{0.948  $\pm$  0.02} \\ [1ex] 
 \midrule
Real noisy with & Corrupted & 21.70 $\pm$  2.15 & 0.328 $\pm$  0.07 & 0.731 $\pm$ 0.05 & 0.094 $\pm$ 0.02\\  
severe harmonic & GradMC & 21.82  $\pm$  2.19 & 0.332  $\pm$  0.08 & 0.096  $\pm$  0.03 & 0.730  $\pm$  0.06\\ 

 &U-Net &23.93  $\pm$  2.18 & 0.441  $\pm$  0.08 & 0.144  $\pm$  0.02 & 0.836  $\pm$  0.03 \\
 
 &Autofocusing & 25.17 $\pm$  2.21 & 0.491  $\pm$  0.11 & 0.219  $\pm$  0.05 & 0.899  $\pm$  0.04 \\
 &\textbf{Autofocusing+(Ours)} & \textbf{26.45  $\pm$  2.19} & \textbf{0.579  $\pm$  0.08} & \textbf{0.261  $\pm$  0.04} & \textbf{0.933  $\pm$  0.02} \\ [1ex] 
 \bottomrule \label{s:all_metrics}
\end{tabular}
\end{table}

\begin{figure}
\includegraphics[width=0.9\textwidth]{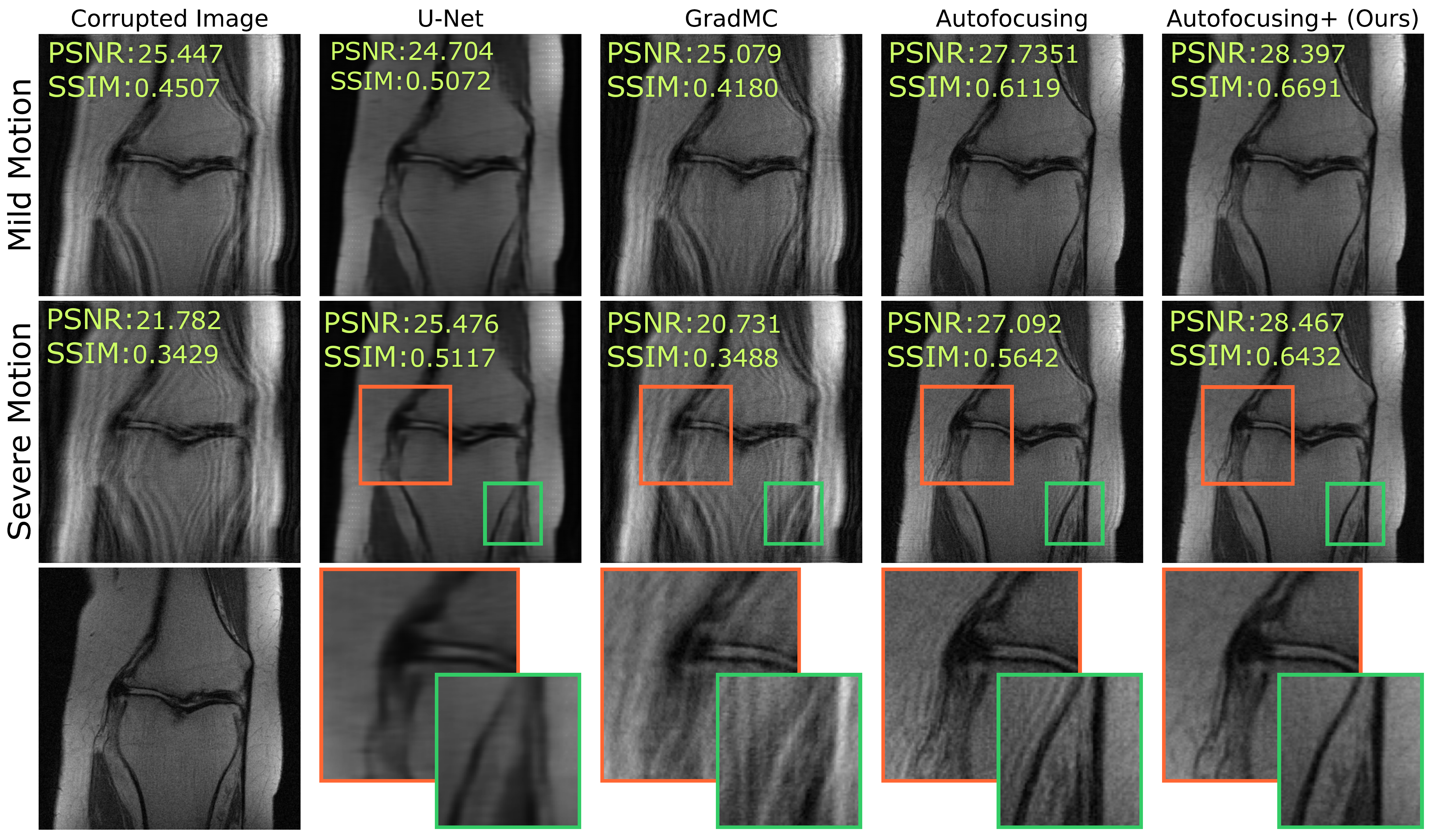}
\caption{Examples of the demotion performance of different methods.} \label{fig:restored_img_2}
\end{figure}

% \end{document}
\end{document}